\documentclass[a4paper, reqno, 11pt]{amsart}

\usepackage[left=2.5cm, right=2.5cm, top=3cm, bottom=3cm]{geometry}

\usepackage{amsaddr}
\usepackage{amssymb}
\usepackage{amsfonts}
\usepackage{mathrsfs}
\usepackage[T1]{fontenc}
\usepackage[USenglish]{babel}
\usepackage{varioref}
\usepackage{multirow}
\usepackage{array,multicol}
\usepackage{bigdelim}
\usepackage{bigstrut}
\usepackage{bm}
\usepackage{units}
\usepackage{bbold}
\usepackage[font=small, labelfont=bf]{caption}
\usepackage[list=true, labelformat=brace, font=small, labelfont=bf]{subcaption}
\usepackage{graphicx}
\usepackage{csquotes}
\usepackage{longtable}
\usepackage{here}

\usepackage[breaklinks=true
]{hyperref}
\hypersetup{pdftitle={F.Nill: On the redundancy of birth and death rates in homogenous epidemic SIR models.}}

\usepackage[backend=biber,style=numeric]{biblatex}

\usepackage{enumitem}
\setlist[itemize]{align=parleft,left=0pt..1.5em}
\setlist{noitemsep}

\usepackage{url}

\def\customauthor{\empty}
\def\customdate{\empty}
\let\oldauthor\author
\renewcommand{\author}[1]{\def\customauthor{#1}}
\renewcommand{\date}[1]{\def\customdate{#1}}
\oldauthor{\customauthor\\\customdate}

\theoremstyle{definition}
\newtheorem{definition}{Definition}[section]

\theoremstyle{plain}

\newtheorem{lemma}[definition]{Lemma}
\newtheorem{corollary}[definition]{Corollary}
\newtheorem{proposition}[definition]{Proposition}

\theoremstyle{remark}
\newtheorem{remark}[definition]{Remark}

\numberwithin{equation}{section}

\newcommand{\ncm}{\newcommand}
\ncm{\rncm}{\renewcommand}

\ncm{\lb}[1]{\label{#1}}

\rncm{\sec}{\setc{0}\section}
\ncm{\bsn}{\bigskip\noindent}
\ncm{\msn}{\medskip\noindent}
\ncm{\ssn}{\smallskip\noindent}
\ncm{\beq}{\begin{equation}}
\ncm{\beqnon}{\begin{equation*}}
\ncm{\eeq}{\end{equation}}
\ncm{\eeqnon}{\end{equation*}}
\ncm{\bea}{\begin{eqnarray}}
\ncm{\beanon}{\begin{eqnarray*}}
\ncm{\eea}{\end{eqnarray}}
\ncm{\eeanon}{\end{eqnarray*}}

\ncm{\ba}{\begin{array}}
\ncm{\ea}{\end{array}}

\ncm{\fns}{\footnotesize}

\DeclareMathOperator\const{const.}

\DeclareMathOperator\age{age}

\ncm{\scenA}{\ensuremath{\mathrm{(A)}}}
\ncm{\scenB}{\ensuremath{\mathrm{(B)}}}
\ncm{\scenC}{\ensuremath{\mathrm{(C)}}}
\ncm{\scenD}{\ensuremath{\mathrm{(D)}}}
\ncm{\scenE}{\ensuremath{\mathrm{(E)}}}

\ncm{\scenAp}{\ensuremath{\mathrm{(A^+)}}}
\ncm{\scenBp}{\ensuremath{\mathrm{(B^+)}}}
\ncm{\scenCp}{\ensuremath{\mathrm{(C^+)}}}
\ncm{\scenDp}{\ensuremath{\mathrm{(D^+)}}}
\ncm{\scenEp}{\ensuremath{\mathrm{(E^+)}}}

\ncm{\scenAm}{\ensuremath{\mathrm{(A^-)}}}
\ncm{\scenBm}{\ensuremath{\mathrm{(B^-)}}}
\ncm{\scenCm}{\ensuremath{\mathrm{(C^-)}}}
\ncm{\scenDm}{\ensuremath{\mathrm{(D^-)}}}

\ncm{\scenXpm}{\ensuremath{\mathrm{(X_\pm)}}}

\newcommand{\RR}{\ensuremath{\mathbb{R}}}
\newcommand{\RRN}{\ensuremath{\mathbb{R}_{\geq 0}}}
\ncm{\NN}{\ensuremath{\mathbb{N}}}
\ncm{\ZZ}{\ensuremath{\mathbb{Z}}}
\ncm{\GG}{\ensuremath{\mathbb{G}}}
\rncm{\SS}{\mathbb{S}}
\ncm{\bone}{\ensuremath{\mathbb{1}}}
\newcommand{\II}{{ \mathbb{I}}}
\newcommand{\bA}{ \mathbf{A}}
\newcommand{\bB}{ \mathbf{B}}
\newcommand{\bC}{ \mathbf{C}}

\newcommand{\bF}{ \mathbf{F}}

\newcommand{\bp}{ \mathbf{p}}

\ncm{\pt}{\bp_\tau}
\newcommand{\bfa}{ \mathbf{a}}

\newcommand{\bzero}{\mathbf{0}}

\newcommand{\A}{\ensuremath{{\mathcal A}}}

\newcommand{\B}{\ensuremath{{\mathcal B}}}

\newcommand{\C}{\ensuremath{{\mathcal C}}}
\newcommand{\D}{\ensuremath{{\mathcal D}}}

\newcommand{\G}{\ensuremath{{\mathcal G}}}
\newcommand{\K}{\ensuremath{{\mathcal K}}}

\newcommand{\V}{\ensuremath{{\mathcal V}}}
\newcommand{\T}{\ensuremath{{\mathcal T}}}

\rncm{\P}{\ensuremath{{\mathcal P}}}
\ncm{\PPe}{\P_\epsilon}

\rncm{\S}{\ensuremath{{\mathcal S}}}

\renewcommand{\L}{\ensuremath{{\mathcal L}}}
\ncm{\Labc}{\L_{a,b,c}}
\newcommand{\I}{\ensuremath{{\mathcal I}}}

\newcommand{\M}{\ensuremath{{\mathcal M}}}

\rncm{\O}{\mathcal{O}}

\ncm{\Me}{\M_\epsilon}

\ncm{\Pph}{\P_{\mathrm{phys}}}
\ncm{\Tph}{\T_{\mathrm{phys}}}
\ncm{\bTph}{\bar{\T}_{\mathrm{phys}}}
\ncm{\Iend}{\I_{\mathrm{end}}}
\ncm{\Ta}{\T_{\bm{a}}}
\ncm{\Aph}{\A_{\mathrm{phys}}}
\ncm{\Abio}{\A_{\mathrm{bio}}}
\ncm{\Asoc}{\A_{\mathrm{soc}}}
\ncm{\Csoc}{\C_{\mathrm{soc}}}
\ncm{\Dsoc}{\D_{\mathrm{soc}}}
\ncm{\Asig}{\A_{\mathrm{split}}}
\ncm{\Csig}{\C_{\mathrm{split}}}
\ncm{\Dsig}{\D_{\mathrm{split}}}
\ncm{\Aone}{\A_{\mathrm{Model-1}}}
\ncm{\Atwo}{\A_{\mathrm{Model-2}}}
\ncm{\Asirs}{\A_{\mathrm{SIRS}}}
\ncm{\Asiso}{\A_{\mathrm{SIS_1}}}
\ncm{\Asist}{\A_{\mathrm{SIS_2}}}
\ncm{\Asisp}{\A_{\mathrm{SIS+}}}
\ncm{\Asism}{\A_{\mathrm{SIS-}}}
\ncm{\Asispm}{\A_{\mathrm{SIS\pm}}}
\ncm{\Dsis}{\D_{\mathrm{SIS}}}
\ncm{\Dsisp}{\D_{\mathrm{SIS+}}}
\ncm{\Dsism}{\D_{\mathrm{SIS-}}}
\ncm{\Dsispm}{\D_{\mathrm{SIS\pm}}}
\ncm{\Asisj}{\A_{\mathrm{SIS_j}}}
\ncm{\Aheth}{\A_{\mathrm{Heth}}}
\ncm{\Dheth}{\D_{\mathrm{Heth}}}
\ncm{\Kheth}{\K_{\mathrm{Heth}}}
\ncm{\Cph}{\C_{\mathrm{phys}}}
\ncm{\Cbio}{\C_{\mathrm{bio}}}
\ncm{\Cone}{\C_{\mathrm{Model-1}}}
\ncm{\Ctwo}{\C_{\mathrm{Model-2}}}
\ncm{\Csirs}{\C_{\mathrm{SIRS}}}
\ncm{\Dph}{\D_{\mathrm{phys}}}
\ncm{\bDph}{\bar{D}_{\mathrm{phys}}}
\ncm{\DBA}{\D_{AB}}
\ncm{\DAB}{\D_{AB}}
\ncm{\Dbio}{\D_{\mathrm{bio}}}
\ncm{\Dbionu}{\D_{{\mathrm{bio},\nu}}}
\ncm{\Dbioz}{\D_{{\mathrm{bio},0}}}
\ncm{\Done}{\D_{\mathrm{Model-1}}}
\ncm{\Dtwo}{\D_{\mathrm{Model-2}}}
\ncm{\DII}{\D_{\RN{2}}}
\ncm{\Dsirs}{\D_{\mathrm{SIRS}}}
\ncm{\Ksirs}{\K_{\mathrm{SIRS}}}
\ncm{\Kbio}{\K_{\mathrm{bio}}}
\ncm{\Kbionu}{\K_{{\mathrm{bio},\nu}}}
\ncm{\Lbio}{\L_{\mathrm{bio}}}
\ncm{\Gdil}{G_{\mathrm{dil}}}
\ncm{\GX}{G_{X}}
\ncm{\GI}{G_{I}}
\ncm{\GS}{G_{S}}

\ncm{\tAph}{\tilde{\A}_{\mathrm{phys}}}
\ncm{\tAbio}{\tilde{\A}_{\mathrm{bio}}}
\ncm{\tAone}{\tilde{\A}_{\mathrm{Model-1}}}
\ncm{\tAtwo}{\tilde{\A}_{\mathrm{Model-2}}}
\ncm{\tAsirs}{\tilde{\A}_{\mathrm{SIRS}}}
\ncm{\TABC}{T(\bA,\bB,\bC)}
\ncm{\Tabc}{\TABC^{\leq 1}}

\ncm{\yph}{y_{\mathrm{phys}}}
\ncm{\Padm}{P_{\mathrm{adm}}}
\ncm{\Plow}{P_{\mathrm{low}}}
\ncm{\Phigh}{P_{\mathrm{high}}}
\ncm{\plow}{p_{\mathrm{low}}}
\ncm{\phigh}{p_{\mathrm{high}}}
\ncm{\Pc}{\P_{\mathrm{cut}}}
\ncm{\Reff}{X_{\mathrm{rep}}}
\ncm{\Reffo}{X_{\mathrm{rep,0}}^*}
\ncm{\Reffe}{X_{\mathrm{rep,end}}^*}
\ncm{\Ie}{I_{\mathrm{end}}^*}
\ncm{\Se}{I_{\mathrm{end}}^*}
\ncm{\dReff}{\dot{X}_{\mathrm{rep}}}
\ncm{\Tosc}{T_{\mathrm{osc}}}
\ncm{\Timm}{T_{\mathrm{imm}}}
\ncm{\Tinf}{T_{\mathrm{inf}}}
\ncm{\Thalf}{T_{\mathrm{half}}}

\ncm{\rvac}{r_{\mathrm{vac}}}

\newcommand{\al}{\alpha}

\newcommand{\be}{\beta}

\newcommand{\ga}{\gamma}

\newcommand{\Del}{\Delta}
\newcommand{\del}{\delta}

\newcommand{\ep}{\epsilon}
\newcommand{\vep}{\varepsilon}
\newcommand{\om}{\omega}

\ncm{\OP}{\Omega_{\PP,\ep}}
\ncm{\oG}{\omega_\G}
\ncm{\ome}{\omega_\ep}
\ncm{\phit}{\varphi_\tau}
\ncm{\p}{\psi}

\ncm{\Aal}{A_\alpha}
\ncm{\Bal}{B_\alpha}
\ncm{\sal}{\sigma_\alpha}

\rncm{\k}{\kappa}
\ncm{\an}{a_\nu}
\newcommand{\id}{{\rm id}}

\newcommand{\bra}{\langle}
\newcommand{\ket}{\rangle}
\def\cros{\,\raise1.9pt\hbox{$\scriptscriptstyle  > $}\!
          \raise1.5pt\hbox{$\scriptstyle\triangleleft$}\,}
\def\>cros{\cros}
\def\<cros{\,\raise1.5pt\hbox{$\scriptstyle\triangleright$}\!
           \raise1.9pt\hbox{$\scriptscriptstyle < $}\,}
\ncm{\veq}{{\scriptstyle\Vert}}

\ncm{\dR}{\partial_R}
\ncm{\dS}{\partial_S}
\ncm{\dI}{\partial_I}
\ncm{\dD}{\partial_D}
\ncm{\dN}{\partial_N}
\ncm{\dM}{\partial_M}
\ncm{\dX}{\partial_X}
\ncm{\dq}{\partial_q}
\ncm{\dx}{\partial_x}
\ncm{\dy}{\partial_y}
\ncm{\parH}{\partial H}
\ncm{\parHe}{\partial H_{\epsilon}}
\ncm{\parq}{\partial q}
\ncm{\parp}{\partial p}

\ncm{\rto}{\rightarrow}
\ncm{\mto}{\longmapsto}
\ncm{\lto}{\longrightarrow}
\ncm{\Lto}{\Longrightarrow}
\ncm{\lra}{\leftrightarrow}
\ncm{\LRA}{\Leftrightarrow}
\ncm{\LLRA}{\Longleftrightarrow}
\ncm{\LRa}{\Leftrightarrow}
\ncm{\LLRa}{\Longleftrightarrow}

\ncm{\tOP}{\tilde{\Omega}_{\PP,\ep}}
\ncm{\toe}{\tilde{\omega}_\ep}
\ncm{\tHe}{\tilde{H}_{\epsilon}}
\ncm{\tH}{\tilde{H}}
\ncm{\tV}{\tilde{V}}
\ncm{\tK}{\tilde{K}}
\ncm{\tE}{\tilde{E}}
\ncm{\tA}{\tilde{\A}}
\ncm{\tP}{\tilde{\P}}
\ncm{\tbF}{\tilde{\bF}}
\ncm{\rt}{\tilde{r}_0}
\ncm{\tr}{\tilde{r}_0}
\ncm{\tga}{\tilde{\gamma}}
\ncm{\tal}{\tilde{\alpha}}
\ncm{\tGa}{\tilde{\Gamma}}
\ncm{\Nt}{\tilde{N}}
\ncm{\tHPe}{\tilde{H}_{\PP,\epsilon}}
\ncm{\tre}{\tilde{\rho}_{\epsilon}}
\ncm{\tq}{\tilde{q}}
\ncm{\tp}{\tilde{p}}
\ncm{\ta}{\tilde{a}}
\ncm{\tb}{\tilde{b}}
\ncm{\tbe}{\tilde{\beta}}
\ncm{\tc}{\tilde{c}}
\ncm{\td}{\tilde{d}}
\ncm{\tep}{\tilde{\vep}}
\ncm{\etq}{e^{\tilde{q}}}
\ncm{\etp}{e^{\tilde{p}}}
\ncm{\ttau}{\tilde{\tau}}

\ncm{\hH}{\hat{H}}
\ncm{\hV}{\hat{V}}
\ncm{\hK}{\hat{K}}
\ncm{\hKs}{\hat{K}_\sigma}
\ncm{\hHs}{\hat{H}_\sigma}
\ncm{\hvc}{\hat{v}_C}
\ncm{\hP}{\hat{\P}}
\ncm{\hA}{\hat{\A}}
\ncm{\hAph}{\hat{\A}_{\mathrm{phys}}}
\ncm{\hB}{\hat{\B}}
\ncm{\hC}{\hat{\C}}
\ncm{\hCph}{\hat{\C}_{\mathrm{phys}}}
\ncm{\hD}{\hat{\D}}
\ncm{\hphi}{\hat{\phi}}
\ncm{\crho}{\check{\rho}}
\ncm{\Reflat}{R_1^{\ \flat}}

\ncm{\ok}{\checkmark}
\ncm{\no}{\ding{55}}
\ncm{\s}{\mathsf{s}}
\ncm{\g}{\mathsf{g}}
\ncm{\h}{\mathsf{h}}
\ncm{\HIG}{H_\alpha}
\ncm{\Hal}{H_\alpha}
\ncm{\oIG}{\omega_\alpha}
\ncm{\HPLV}{H_{\mathrm {pLV}}}
\ncm{\omPLV}{\om_{\mathrm {pLV}}}
\ncm{\Hreg}{H^{\mathrm {reg}}}
\ncm{\omreg}{\om^{\mathrm {reg}}}
\ncm{\HPLVreg}{H_{\mathrm {pLV}}^{\mathrm {reg}}}
\ncm{\omPLVreg}{\omega_{\mathrm {pLV}}^{\mathrm{reg}}}
\ncm{\Halreg}{H_{\alpha}^{\mathrm {reg}}}
\ncm{\omalreg}{\omega_{\alpha}^{\mathrm {reg}}}
\ncm{\xnull}{x_{\mathrm {null}}}
\ncm{\Lt}{L_\tau}
\ncm{\Lmin}{L_{\min}}
\ncm{\dLt}{\dot{L}_\tau}
\ncm{\rv}{a_\mathrm{vac}}

\ncm{\DBo}{\D_{\Del B>0}}
\ncm{\DelB}{\Del B}

\newcommand{\Eqref}[1]{Eq. \eqref{#1}}

\ncm{\ulim}[1]{\underset{#1}{\lim}}
\ncm{\secref}[1]{Section \ref{#1}}
\ncm{\figref}[1]{Fig. \ref{#1}}

\newcommand{\0}{_{(0)}}

\newcommand{\minus}{\scalebox{0.75}[1.0]{$-$}}
\newcommand{\inv}{^{\minus 1}}
\ncm{\vsir}{V_{SIR}}
\ncm{\Vsir}{V_{SIR}}
\ncm{\hsir}{H_{SIR}}

\ncm{\zit}[1]{\autocite{#1}}
\ncm{\GHS}{\textsc{HGS }}
\ncm{\Upot}{$U\!$-potential}
\ncm{\QUpot}{quasi-\Upot}
\ncm{\UE}{{V^E}}
\ncm{\VE}{{V^E}}
\ncm{\alpm}{\upsilon_\pm}
\ncm{\alp}{\upsilon_+}
\ncm{\alm}{\upsilon_-}
\ncm{\vpm}{\upsilon_\pm}
\ncm{\xpm}{x_\pm}
\ncm{\qpm}{q_\pm}

\ncm{\vp}{\upsilon_+}
\ncm{\vm}{\upsilon_-}
\ncm{\vO}{v_{\O}}
\ncm{\vN}{v_N}
\ncm{\vt}{v_\tau}
\ncm{\ut}{u_\tau}
\ncm{\apm}{a_\pm}
\ncm{\bpm}{b_\pm}
\ncm{\upm}{u_\pm}
\ncm{\qp}{q_+}
\ncm{\qc}{q_c}
\ncm{\up}{u_+}
\ncm{\xp}{x_+}
\ncm{\xc}{x_c}
\ncm{\epm}{\varepsilon_\pm}
\ncm{\fpm}{f_\pm}
\ncm{\Apm}{A_\pm}
\ncm{\Bpm}{B_\pm}
\ncm{\Dpm}{D_\pm}
\ncm{\Dp}{\vp}
\ncm{\Dc}{\Delta_c}

\ncm{\Spm}{S_\pm}
\ncm{\rSpm}{\rho S_\pm}
\ncm{\thpm}{\theta_\pm}

\ncm{\ntg}{\notag\\}

\ncm{\Ss}{S_{\textsl{sample}}}
\ncm{\Is}{I_{\textsl{sample}}}
\ncm{\Zs}{Z_{\textsl{sample}}}
\ncm{\Es}{E_{\textsl{sample}}}
\ncm{\Ns}{N_{\textsl{sample}}}
\ncm{\rhos}{\rho_{\textsl{sample}}}
\ncm{\gs}{\gamma_{\textsl{sample}}}
\ncm{\Zrge}{Z_{\rho,\gamma,E}}
\ncm{\Zmax}{Z_{\max}}
\ncm{\Qmax}{Q_{\max}}
\ncm{\el}{e^{\lambda}}
\ncm{\Ve}{V_{\epsilon}}
\ncm{\He}{H_{\epsilon}}

\ncm{\HPe}{H_{\PP,\epsilon}}
\ncm{\Emax}{E_{\max}}
\ncm{\rep}{\rho_{\epsilon}}
\ncm{\qe}{q_{\epsilon}}
\ncm{\expe}{\exp_{\epsilon}}
\ncm{\lne}{\ln_{\epsilon}}
\ncm{\Vpme}{V_{\pm,\ep}}
\ncm{\Vpe}{V_{+,\ep}}
\ncm{\Vme}{V_{-,\ep}}
\ncm{\qpme}{q_{\pm,\ep}}
\ncm{\qpe}{q_{+,\ep}}
\ncm{\qme}{q_{-,\ep}}
\ncm{\xpme}{x_{\pm,\ep}}
\ncm{\xpe}{x_{+,\ep}}
\ncm{\xme}{x_{-,\ep}}
\ncm{\qG}{q_\G}
\ncm{\pG}{p_\G}
\ncm{\ys}{y_2^*}
\ncm{\xs}{x_1^*}
\ncm{\vs}{v_2^*}
\ncm{\us}{u_1^*}
\ncm{\fl}{\varphi_\tau}
\ncm{\Gas}{\Gamma_\sigma}
\ncm{\tmp}{\age}
\ncm{\tImp}{\tau_{\I,\max}}

\begin{document}

\title[Redundancy of birth and death rates]
{On the redundancy of birth and death rates in homogenous epidemic SIR models}
\author{Florian Nill}


\email{florian.nill@fu-berlin.de}



\address{%
Fachbereich Physik, Freie Universität Berlin, 14195 Berlin, Germany}

\begin{abstract}
The dynamics of fractional population sizes $y_i=Y_i/N$ in homogeneous compartment models with time dependent total population $N$ is analyzed. Assuming constant per capita birth and death rates the vector field $\dot{Y}_i=V_i(Y)$ naturally projects to a vector field $F_i(Y)$ tangent to the leaves of constant population $N$. A universal formula for the projected field $F_i$ is given. In this way, in many SIR-type models with standard incidence all demographic parameters become redundant for the dynamical system 
$\dot{y}_i=F_i(y)$. They may be put to zero by shifting remaining parameters appropriately. Normalizing eight examples from the
literature this way, they unexpectedly become isomorphic for corresponding parameter ranges. Thus, some recently published results turn out to be already covered by papers 20 years ago.
\end{abstract}
%

\subjclass{34C23, 34C26, 37C25, 92D30}
\keywords{SIRS model, demographic parameters, birth and10195 death rates, normalization.}
\maketitle




\section{Introduction}
\begin{figure}[ht!]
	  \captionsetup{singlelinecheck=true}
\centering
\includegraphics[width=0.5\textwidth]{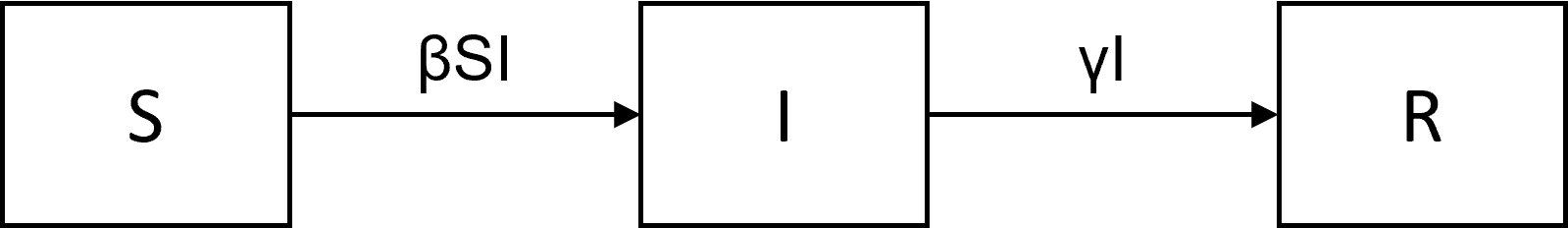}
	\caption{Flow diagram of the SIR model.}
	\label{Fig_SIR}
\end{figure}

The classic SIR model had been introduced by Kermack and McKendrick in 1927 \cite{KerMcKen} as one of the first models in mathematical epidemiology. The model divides a population into three compartments with fractional sizes $S$ (Susceptibles), $I$ (Infectious) and $R$ (Recovered), such that $S+I+R=1$. The flow diagram between compartments as given in Fig. \ref{Fig_SIR} leads to the dynamical system
\begin{equation}
\dot{S}=-\be SI,\qquad\dot{I}=\be SI-\ga I,\qquad\dot{R}=\ga I.
\label{SIR}
\end{equation}

Here $\ga$ denotes the recovery 
rate and $\be$ the effective contact rate (i.e. the number of contacts/time of a susceptible leading to an infection given the contacted was infectious). Members of $R$ are supposed to be immune forever. By \eqref{SIR} $S$ decreases monotonically causing eventually  
$\be S<\ga$ and $\dot{I}<0$. At the end the disease dies out, $I(\infty)=0$, and one stays with a nonzero final size 
$S(\infty)>0$, thus providing a model for {\em Herd immunity}. 

\begin{figure}[ht!]
\captionsetup{singlelinecheck=true}
\centering
\begin{subfigure}[b][][b]{0.4\linewidth}
\centering
\includegraphics[width=\textwidth]{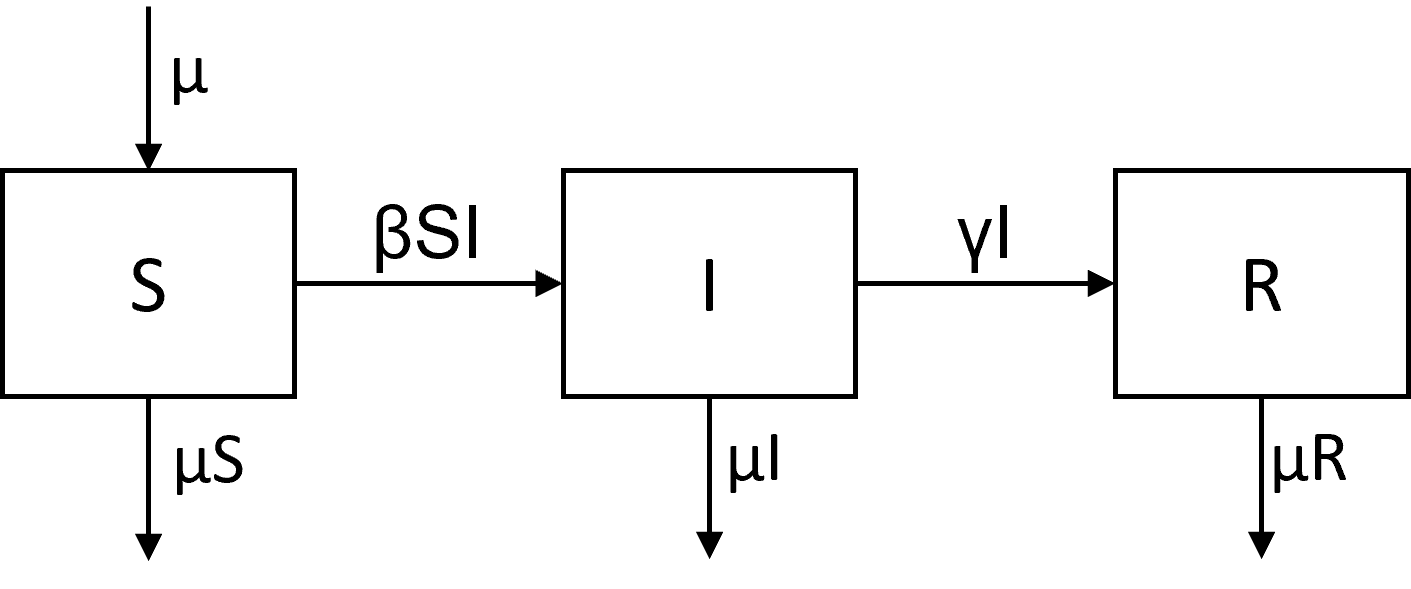}
	\caption{Hethcote's model}
\end{subfigure}

\begin{subfigure}[b][][b]{0.4\linewidth}
\includegraphics[width=\textwidth]{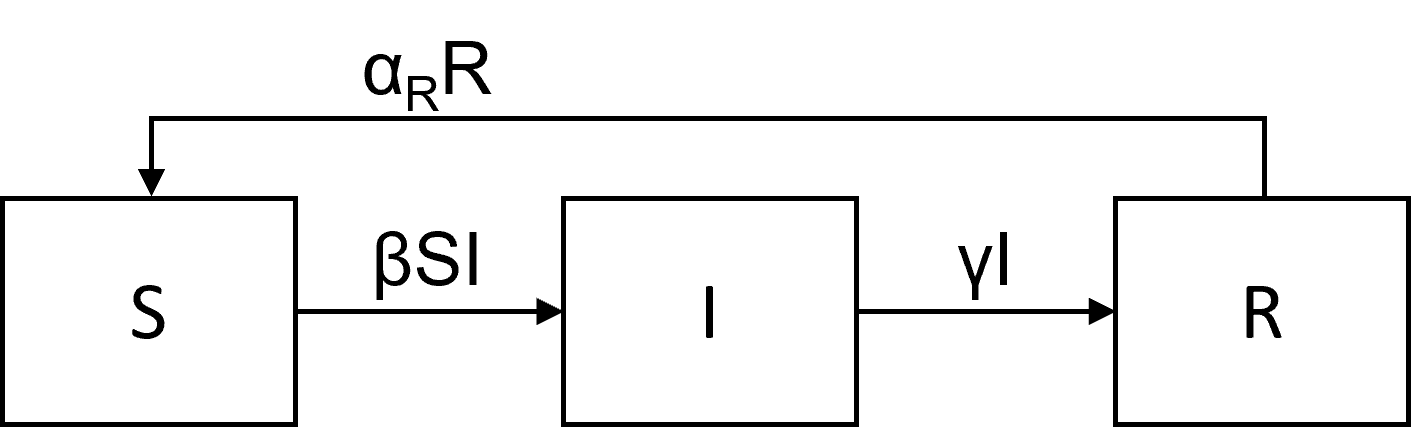}
	\caption{SIRS model}
\end{subfigure}
\qquad
\begin{subfigure}[b][][b]{0.25\linewidth}
\includegraphics[width=\textwidth]{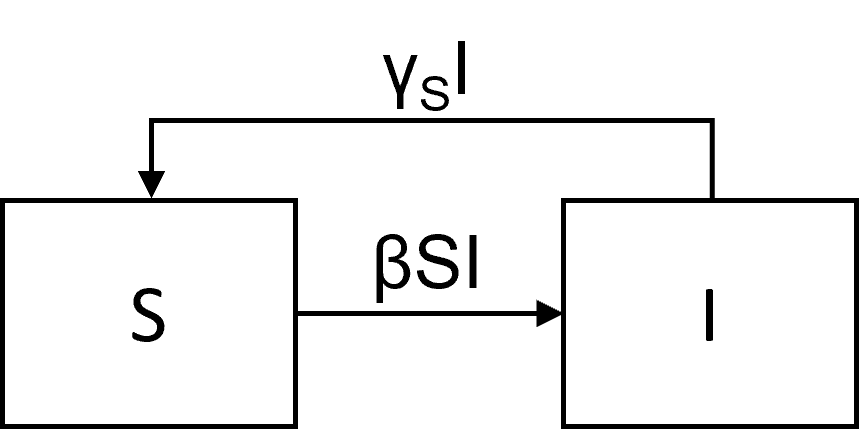}
	\caption{SIS model}
\end{subfigure}
\caption{Standard models featuring endemic equilibria}
	\label{Fig_endemic}
\end{figure}

To construct models featuring also {\em endemic} scenarios one needs enough supply of susceptibles to keep the incidence $\be SI$ ongoing above a positive threshold. The literature discusses three basic methods to achieve this, see Fig. \ref{Fig_endemic}.

\begin{itemize}
\item
Hethcotes {\em classic endemic model} adds to the SIR model  balanced birth and death rates $\mu$  and assumes all newborns susceptible. This leads to a 
bifurcation from a stable disease-free equilibrium point to a stable endemic
scenario when raising the basic reproduction number $r_0=\be/\ga$ above one \cite{Hethcote1974,Hethcote1976, Hethcote1989}.
\item
The {\em SIRS model} adds to the SIR model an immunity waning flow
$\al_R R$ from $R$ to $S$, leading to the same result.
\item
The {\em SIS model} considers recovery without immunity, i.e. a recovery flow $\ga_S I$ from $I$ to $S$ while putting $R=0$. Again this leads to the same result.
\end{itemize}

In what follows the reader is assumed to be familiar with the basic notions in these models.
For a comprehensive and self-contained overview of history, methods and results on mathematical epidemiology see the textbook by M. Martcheva, \cite{Martcheva}, where also an extensive list of references to original papers is given.

\begin{figure}[ht!]
\captionsetup{singlelinecheck=true}
\centering
\begin{subfigure}[c][][b]{0.4\linewidth}
\centering
\includegraphics[width=\textwidth]{Flow_Hethcote.png}
	\caption{Hethcote's model}
\end{subfigure}
{\LARGE $\ \ \bm{\cong}\ \ $}
\begin{subfigure}[c][][b]{0.4\linewidth}
\includegraphics[width=\textwidth]{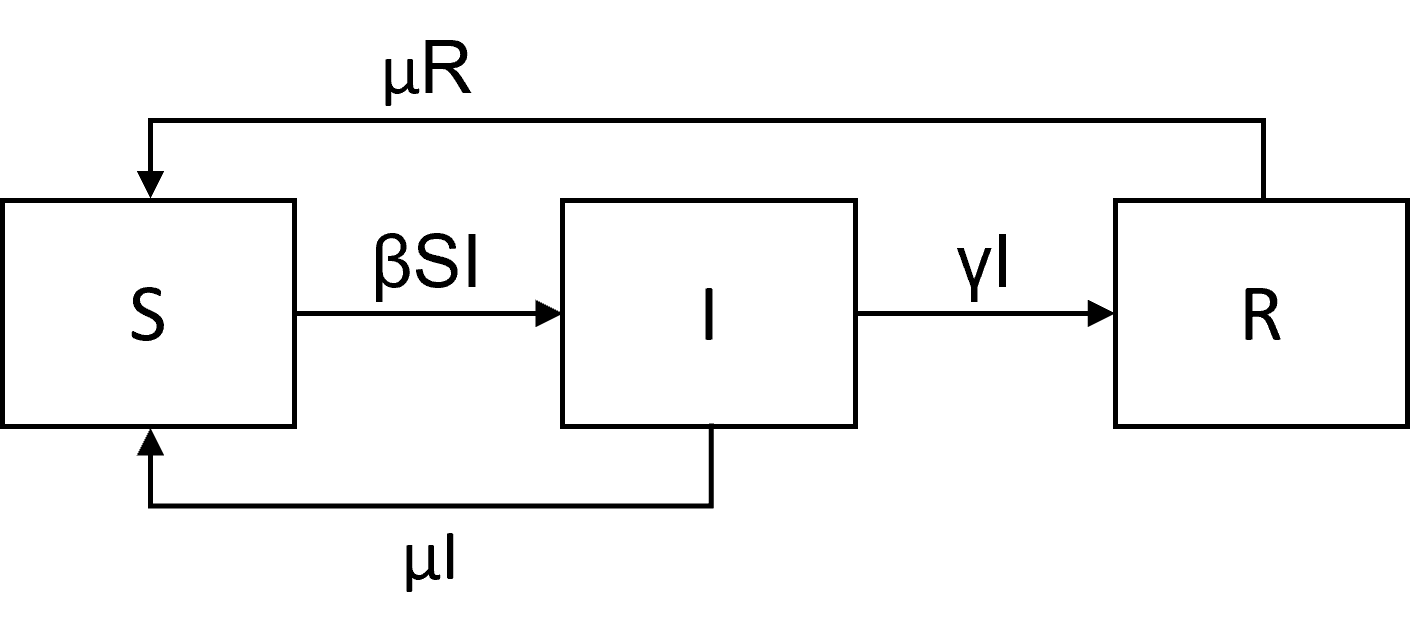}
	\caption{SI(R)S model}
\end{subfigure}
\caption{Equivalence of models using $\mu-\mu S=\mu I+\mu R$}
	\label{Fig_heth}
\end{figure}

\bsn
As a starting point for this paper observe from Fig. \ref{Fig_heth} that Hethcote's model could equivalently be reformulated by disregarding birth and death rates and instead introducing  
a combined SI(R)S $\equiv$ SIRS/SIS model with flow rates 
$\ga_S=\al_R=\mu$.
More generally, adding Hethcote's balanced birth and death rates 
$\mu$ to a SI(R)S model with independent parameters $(\ga_S,\al_R)$ apparently becomes equivalent  to considering the SI(R)S model without birth and death rates and with shifted parameters 
$\tilde{\ga}_S=\ga_S+\mu$ and $\tilde{\al}_R=\al_R+\mu$
\cite{Nill_Omicron}.

The aim of this letter is to generalize this observation to homogeneous 3-compartment models with

A) positive susceptibility of the $R$-compartment describing incomplete immunity (in which case it makes sense to rename 
$S\equiv S_1$ and $R\equiv S_2$),

B) a non-trivial birth matrix and a time varying population size $N$ due to compartment dependent constant per capita birth and death rates.

As a result we will see that for coinciding birth-minus-death rates in compartments $S_1\equiv S$ and $S_2\equiv R$ in the dynamics of fractional variables all demographic parameters become redundant by shifting remaining parameters appropriately. 
In particular, transmission coefficients $\be_i$ describing $S_i$-susceptibility are replaced by 
$\tbe_i=\be_i-\Del\mu_I$, where 
$\Del\mu_I$ denotes the excess mortality in compartment $I$. Hence, $\tilde{\be_i}$ may possibly become negative. 

This result leads to a unifying normalization prescription by always considering these models 
without vital dynamics and, instead, with two distinguished and possibly also negative incidence rates $\tilde{\be}_i\in\RR$.
When normalized this way, seemingly different models in the literature become isomorphic at coinciding shifted parameters. As an example, recent results of \cite{AvramAdenane2022} already follow from earlier results of \cite{KribsVel} (for $\tbe_2>0$) and \cite{LiMa2002} (for $\tbe_2<0$).

\section{Compartment models\label{Sec_comp-models}}
For simplicity, all maps are supposed to be
$\C^{\infty}$. Let $\V=\RR^n$ and $V:\V\rto\V$ a homogeneous vector field, $V(\lambda Y)=\lambda V(Y)$ for all 
$\lambda\in\RR_+$ and $Y\in\V$. Denote $\V^*$ the dual of $\V$ and $\bra\cdot|\cdot\ket:\V^*\otimes\V\rto\RR$ the dual pairing. Let 
$\phi_t:\V\rto\V$ be the local flow of $V$. For functions 
$f:\V\rto\RR$ we denote their time derivative along $\phi_t$ by 
$\dot{f}:=d/dt|_{t=0} (f\circ\phi_t)=\bra \nabla f| V\ket$. Let 
$0\not\in\P\subset\V$ be a cone and 
$N:\P\rto\RR_+$ be a homogeneous function,
$N(\lambda Y)=\lambda N(Y)$, satisfying  $\nabla N\neq 0$ on 
$\P$. In this case the local flow $\phi_t$ naturally projects to a local flow 
$\psi_t$ leaving the leaves $\{N=\const\}$ invariant.
$$
\psi_t(Y):=N(Y)N(\phi_t(Y))\inv\phi_t(Y)
$$
Using $\phi_t(\lambda Y)=\lambda\phi_t(Y)$ one immediately checks 
$$
\psi_0=\id,\qquad\psi_{t+s}=\psi_t\circ\psi_s,\qquad 
N\circ\psi_t = N\,.
$$
The vector field $F:\P\rto\RR^n$ generating $\psi_t$ is given by 
\begin{equation}
F(Y):=V(Y)-\frac{\dot{N}}{N}Y\quad \Longrightarrow\quad
\dot{\psi_t}=F\circ\psi_t\,.
\label{F}
\end{equation}
Clearly, $F$ is also homogeneous and putting $y:=Y/N$ we have 
$\dot{y}=F(y)$.

\bsn
Now let's specialize to compartment models, where $Y_i$ gives the population in compartment $i$, $N(Y):=\sum_i Y_i$ the total  population and $\P:=\RRN^n\setminus\{\bzero\}$. To guarantee $\P$ being forward invariant one also needs 
$Y_i=0\Rightarrow V_i(Y)\geq 0$.

\begin{definition}
The compartment model $\dot{Y}=V(Y)$ is said to have constant per capita demographic rates, iff there exists 
$\nu=(\nu_1,\cdots,\nu_n)\in \V^*$ such that 
$\dot{N}=\nu$, i.e. $\sum_i V_i(Y)=\sum_i\nu_i Y_i$. We call 
$\nu_i$ the total birth-minus-death rate in compartment $i$.
\end{definition}
\bsn
In such models one usually decouples the time development of $N$ and analyzes the dynamics of fractional variables $y=Y/N$, $\dot{y}=F(y)$.
The main observation of this paper states, that in many standard models the correction term $N\inv\dot{N}Y$ in \Eqref{F} can be absorbed by redefining the parameters determining $V$.

\begin{lemma}\label{Lem_Q}
Assume $\dot{N}=\nu$ and denote  
$Q_{ijk}=[\delta_{ij}(\nu_k-\nu_j)+\delta_{ik}(\nu_j-\nu_k)]/2$. Putting  $Q_i(Y):=\sum_{j,k}Q_{ijk}Y_jY_k$ we have
\begin{align}
\frac{\dot{N}}{N}Y_i	&=\nu_i Y_i+\frac{1}{N}Q_i(Y)
\label{Q}
\end{align}
\end{lemma}
\begin{proof}
Use $y_i=1-\sum_{j\neq i}y_j$ and therefore
$\bra\nu|y\ket=\nu_i+\sum_j(\nu_j-\nu_i)y_j$, for all 
$i=1,\cdots,n$.
\end{proof}
\ssn
Let us apply this to vector fields $V$ of the form
\begin{equation}
V_i(Y)=\sum_j L_{ij}Y_j+\sum_j M_{ij}Y_j +
\frac{1}{N}\sum_{j,k}\Lambda_{ijk}Y_jY_k\,,
\label{V}
\end{equation}
where $\Lambda_{ijk}=\Lambda_{ikj}$, 
 $\sum_i M_{ij}=\sum_i \Lambda_{ijk}=0$
and $L_{ij}=B_{ij}-\delta_{ij}\mu_j$. Here $\mu_j\geq 0$ is the mortality rate in compartment $j$, $B_{ij}Y_j\geq 0$ denotes the number of newborns from compartment $j$ landing in compartment $i$, and the parameters $M_{ij}$ and $\Lambda_{ijk}$ determine the population flow from compartment $j$ to $i$, say due to infection transmission, recovery, loss of immunity, vaccination, etc. Thus, 
$\delta_j:=\sum_i B_{ij}$ is the total birth rate in compartment $j$ and $\nu_j=\sum_i L_{ij}=\delta_j-\mu_j$. Also,
forward invariance of the nonnegative orthant $\P$ for zero birthrates requires a) $M$ to be essentially nonnegative, i.e. 
$M_{ij}\geq 0$ for $i\neq j$, whence $M_{jj}=-\sum_{i\neq j}M_{ij}\leq 0$, and b) $\sum_{k\neq i}\Lambda_{ijk}\geq -M_{ij}$ for $i\neq j$. Now put 
\begin{equation}
\tilde{M}_{ij}:=M_{ij}+L_{ij}-\delta_{ij}\nu_j\equiv
M_{ij}+B_{ij}-\delta_{ij}\delta_j\,,\qquad
\tilde{\Lambda}_{ijk}:=\Lambda_{ijk}-Q_{ijk}\,.
\label{tilde_M}
\end{equation}
Using $B_{ij}\geq 0$, $Q_{ijk}=Q_{ikj}$, 
$\sum_i Q_{ijk}=0$ and $Q_{ijk}=0$ for $j\neq i\neq k$, the new parameters $\tilde{M}$ and 
$\tilde{\Lambda}$ have the same properties as 
$M$ and $\Lambda$ and we get 
\begin{equation}
F_i(Y)=\sum_j\tilde{M}_{ij}Y_j+
\frac{1}{N}\sum_{j,k}\tilde{\Lambda}_{ijk}Y_jY_k\,.
\label{F_i}
\end{equation}
Hence, in the dynamics for fractional variables $y=Y/N$ all birth and death rates may be absorbed by redefining 
$M$ and $\Lambda$. Note that standard models typically satisfy 
$\Lambda_{ijj}=0$, which is consistent with $Q_{ijj}=0$. On the other hand, 
$\tilde{\Lambda}_{iik}=\tilde{\Lambda}_{iki}$ might change sign as compared to epidemiological requirements.

\begin{remark}
If the vector field $V$ is of the form \eqref{V} with $\mu_i$ replaced by $\Del\mu_i+f(Y)$ for some function $f$ and constant excess mortalities $\Del\mu_i$, then $V$ is no longer homogeneous but still $\dot{y}=N\inv F(Y)$. In this case \Eqref{Q} still holds with $\nu_i:=\delta_i-\Del\mu_i-f(Y)$. Hence, the function $f$ does not appear in the definition of $\tilde{M}$ and 
$\tilde{\Lambda}$ in \eqref{tilde_M}, implying that $F$ in \Eqref{F_i} is independent of $f$ and still homogeneous, whence 
$\dot{y}=F(y)$.
\label{Remark}
\end{remark}

\section{The 3-compartment master model\label{Sec_SSISS-model} }
\begin{figure}[ht!]
\centering
\includegraphics[width=0.5\textwidth]
	{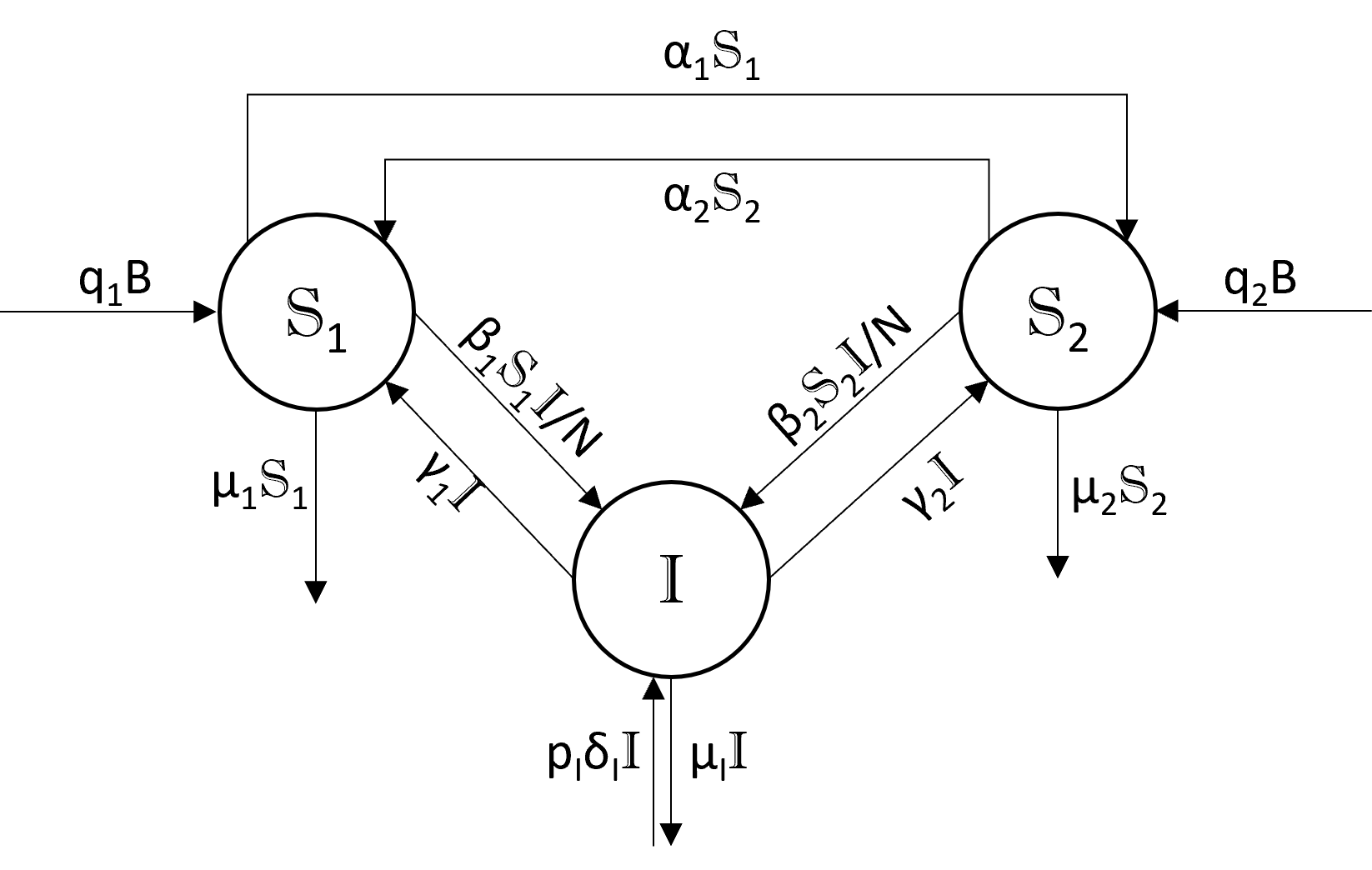}
	\caption{Flow diagram of the master 
	model. 
	$B=\delta_1\mathbb{S}_1+\delta_2\mathbb{S}_2+(1-p_I)\delta_I\II$ denotes the 
	number of not infected newborns per unit of time. 
	}
	\label{Fig_SSI-Flow}
\end{figure}

As a kind of master example consider an abstract  SI(R)S-type model consisting of three compartments, $\mathbb{S}_1$, $\mathbb{S}_2$ and $\mathbb{I}$, with total population $N=\mathbb{S}_1+\mathbb{S}_2+\mathbb{I}$. Members of $\mathbb{I}$ are infectious, members of $\mathbb{S}_1$ are highly susceptible (not immune) and members of $\mathbb{S}_2$ are less susceptible (partly immune). The flow diagram between compartments is completely symmetric with respect to permuting 
$1\leftrightarrow 2$ and depicted in Fig. \ref{Fig_SSI-Flow}.

Parameters in this model are

\smallskip
\begin{tabular}{lcp{0.75\textwidth}}
$\al_1$ &:& 
		Vaccination rate.
\\
$\al_2$ &:& 
		Immunity waning rate.
\\
$\be_i$ &:& 
		Number of effective contacts per unit time of a 
		susceptible from $\mathbb{S}_i$.
\\
$\ga_i$ &:& 
		Recovery rate from $\mathbb{I}\rto \mathbb{S}_i$.
\\
$\mu_i$ &:& 
		Mortality rate in $\mathbb{S}_i$.
\\
$\mu_I$ &:& 
		Mortality rate in $\mathbb{I}$.
\\
$p_I$  &:& 
		Probability of a newborn from $\II$ to be infected.
\\
$\delta_I$ &:& 
		Rate of newborns from $\II$.
\\
$\delta_i$ &:& 
		Rate of newborns from $\mathbb{S}_i$. These are supposed 
		to be not infected.
\\
$B$	&:&
		Sum of not infected newborns, 
		$B=\delta_1\mathbb{S}_1+\delta_2\mathbb{S}_2+(1-p_I)\delta_I\II$. 
\\
$q_i$  &:& 
		Portion of not infected newborns landing in
		$\mathbb{S}_i$, 
		$q_1+q_2=1$. So, $q_2$ is the portion of not infected and 
		vaccinated newborns.
\end{tabular}

\medskip\noindent
All parameters are assumed nonnegative. Also
	$p_I\leq 1$, $q_1+q_2=1$, $\be_1+\be_2>0$ and $\ga_1+\ga_2>0$.
Putting $B:=\delta_1\mathbb{S}_1+\delta_2\mathbb{S}_2+(1-p_I)\delta_I\II$ the dynamics is given by
\begin{align}
\dot{\mathbb{S}}_1 &=q_1 B  + \ga_1\II-
[\mu_1+\al_1+\be_1\II/N]\mathbb{S}_1 +\al_2\mathbb{S}_2
\label{dot_SS1}\\
\dot{\mathbb{S}}_2 &=q_2 B  + \ga_2\II-
[\mu_2+\al_2+\be_2\II/N]\mathbb{S}_2+ \al_1\mathbb{S}_1
\label{dot_SS2}\\
\dot{\II}&=[\be_1\mathbb{S}_1/N +\be_2\mathbb{S}_2/N-\ga_1-\ga_2-\mu_I+p_I\delta_I]\,\II
\label{dot_II}
\end{align}

\ssn
So, in total this model counts 14 independent parameters. A list of prominent examples will be discussed below. Let us now cast this model into the formalism of Section \ref{Sec_comp-models}. 
Putting $Y=(\mathbb{S}_1,\mathbb{S}_2,\II)^T$ and $\Lambda_i(Y)=\sum_{j,k}\Lambda_{ijk}Y_jY_k$ we have
\begin{align}
M	&=
\begin{pmatrix}
-\al_1 & \al_2 & \ga_1
\\
\al_1 & -\al_2 & \ga_2
\\
0 & 0 & -\ga_1-\ga_2
\end{pmatrix}
&
L	&=
\begin{pmatrix}
q_1\delta_1-\mu_1	& q_1\delta_2	& q_1(1-p_I)\delta_I
\\
q_2\delta_1	& q_2\delta_2-\mu_2	& q_2(1-p_I)\delta_I
\\
0	&0	& p_I\del_I-\mu_I
\end{pmatrix}
\\
\Lambda(Y) &=
\begin{pmatrix}
-\be_1\mathbb{S}_1\II
\\
-\be_2\mathbb{S}_2\II
\\
(\be_1\mathbb{S}_1+\be_2\mathbb{S}_2)\II
\end{pmatrix}
&
Q(Y) &=
\begin{pmatrix}
(\nu_I-\nu_1)\mathbb{S}_1\II + (\nu_2-\nu_1)\mathbb{S}_1\mathbb{S}_2
\\
(\nu_I-\nu_2)\mathbb{S}_2\II + (\nu_1-\nu_2)\mathbb{S}_1\mathbb{S}_2
\\
[(\nu_1-\nu_I)\mathbb{S}_1+(\nu_2-\nu_I)\mathbb{S}_2]\II
\end{pmatrix}
\end{align}

\noindent
Here $\nu_i=\del_i-\mu_i$, $\nu_I=\del_I-\mu_I$ and we get
$\dot{N}=\nu_1\mathbb{S}_1+\nu_2\mathbb{S}_2+\nu_I\II$.
So now introduce 
\begin{equation}
\begin{array}{rclrcl}
\tal_1 &:=&\al_1+q_2\delta_1\,,\qquad\qquad\qquad &
\tal_2 &:=&\al_2+q_1\delta_2\,, 
\\
\tga_1 &:=&\ga_1+q_1(1-p_I)\delta_I\,,& 
\tga_2 &:=&\ga_2+q_2(1-p_I)\delta_I\,,
\\
\tilde{\be}_1 &:=&\be_1+\nu_I-\nu_1\,,& 
\tilde{\be}_2 &:=&\be_2+\nu_I-\nu_2\,.
\end{array}
\label{tilde_parameters} 
\end{equation}
to conclude from \eqref{tilde_M}
\begin{align}
\tilde{M}	&=
\begin{pmatrix}
-\tal_1 & \tal_2 & \tga_1
\\
\tal_1 & -\tal_2 & \tga_2
\\
0 & 0 & -\tga_1-\tga_2
\end{pmatrix}
&
\tilde{\Lambda}(Y) &=
\begin{pmatrix}
-\tbe_1\mathbb{S}_1\II
\\
-\tbe_2\mathbb{S}_2\II
\\
(\tbe_1\mathbb{S}_1+\tbe_2\mathbb{S}_2)\II
\end{pmatrix}
+\mathbb{S}_1\mathbb{S}_2
\begin{pmatrix}
\nu_1-\nu_2
\\
\nu_2-\nu_1
\\0
\end{pmatrix}
\end{align}

\bsn
In summary, 
denoting fractions of the total population by $S_i=\mathbb{S}_i/N$ and $I=\mathbb{I}/N$ and assuming the condition 
$\nu_1=\nu_2=:\nu$ the dynamics for fractional variables becomes
\begin{equation}
\begin{pmatrix}
\dot{S}_1
\\
\dot{S}_2
\\
\dot{I}
\end{pmatrix}
=
\begin{pmatrix}
-\tal_1 & \tal_2 & \tga_1
\\
\tal_1 & -\tal_2 & \tga_2
\\
0 & 0 & -\tga_1-\tga_2
\end{pmatrix}
\begin{pmatrix}
S_1\\S_2\\I
\end{pmatrix}
+
\begin{pmatrix}
-\tbe_1S_1 I
\\
-\tbe_2S_2 I
\\
(\tbe_1S_1+\tbe_2S_2)I
\end{pmatrix}
\label{frac_dynamics}
\end{equation}
So, for $\nu_1=\nu_2=\nu$ all birth and death rates become redundant and may be absorbed by redefining 
$\be_i$, $\al_i$ and $\ga_i$. The price to pay is that 
$\tilde{\be}_i=\be_i+\nu_I-\nu$ might become negative.
Hence, the space of admissible parameters for the system \eqref{frac_dynamics} becomes\footnote{The case 
$\tbe_1=\tbe_2$ will be ignored, since in this case putting $S=S_1+S_2$ one easily checks that 
$(S,I)$ obeys the dynamics of a SIS model, which can immediately be solved by separation of variables.  Also, due to the permutation symmetry $1\leftrightarrow 2$, there is no loss assuming $\tbe_1>\tbe_2$.}:
\begin{equation}
\A:=\{(\tilde{\al}_i,\tilde{\beta}_i,\tilde{\ga}_i)\in\RR^6\mid
\tal_i\geq 0,\ \tga_i\geq 0,\ \tga_1+\tga_2>0,\ \tbe_1>\tbe_2\}
\label{A}
\end{equation}
Concerning the dynamics of fractional variables, any two models mapping to the same set of shifted parameters 
$\bfa\in\A$ become isomorphic. In particular, the case of constant population, $\nu_i=\nu_I=0$, yields $\tilde{\be_i}=\be_i$.
In summary we get 

\begin{proposition}\label{Prop_N_varying}
Referring to the parameter  transformation \eqref{tilde_parameters} and the normalized dynamics of fractional variables \eqref{frac_dynamics} assume
$\nu_1=\nu_2=:\nu$ and put $\Del\nu_I:=\nu-\nu_I$.
\begin{itemize}
\item[i)]
If $\Del\nu_I<\be_2$ the model with variable population is isomorphic to a model with constant population and transmission coefficients $\be_i'=\be_i-\Del\nu_I>0$.
\item[ii)]
If $\Del\nu_I>\be_2$ it is isomorphic to a 
variable population SI(R)S model with two recovery flows $I\rto S_1$ and $I\rto S_2$ and parameters
$\be'_2=0$, $\be'_1=\be_1-\be_2$ and 
$\Del\nu'_I=\Del\nu_I-\be_2$.
\item[iii)]
If $\Del\nu_I=\be_2$ it is isomorphic to a SI(R)S model as in ii) with constant population.
\end{itemize} 
\end{proposition}

\section{Examples from the literature\label{Sec_examples}}
For simplicity, from now on let's assume the rate of not infected newborns to be compartment independent, 
$\delta_1=\delta_2=(1-p_I)\delta_I=\delta$, implying $B=\delta N$.
Also, in this case one may without loss assume $p_I=0$ by redefining $\mu_I$. Hence $\nu_1=\nu_2\LRA\mu_1=\mu_2=:\mu$  
and in this case $\Del\nu_I=\Del\mu_I:=\mu_I-\mu$ gives the
{\em excess mortality} in the infectious compartment. 

\bsn
Below there is a list of prominent examples from the literature.
Table \ref{Tab_examples} maps these examples 
to the present set of parameters.

\begin{flushleft}
\begin{longtable}{@{}lp{0.9\textwidth}@{}}
Heth &
Hethcotes classic endemic model 
\cite{Hethcote1974, Hethcote1976, Hethcote1989} by putting $\delta=\mu_i=\mu_I>0$, $q_1=1$, $\beta_1>0$, 
$\ga_2>0$ and all other parameters vanishing.
\\
BuDr	&
The 7-parameter SIRS model with time varying population size in \cite{BusDries90}, adding to Hethcote's model  an immunity waning rate $\al_2$ and allowing non-balancing mortality and birth rates $\delta\neq\mu_i\neq\mu_I$.
\\
SIRI	&
The 6-parameter SIRI model of \cite{DerrickDriessche}, replacing the immunity waning rate $\al_2$ in \cite{BusDries90} by the transmission rate $\be_2>0$ and also requiring $\mu_1=\mu_2$.
\\
SIRS	&
The 8-parameter constant population SI(R)S model with vaccination and two recovery flows $I\rightarrow S_1$ and 
$I\rightarrow S_1$. Hence 
$\delta=\mu_i=\mu_I$ and $\beta_2=0$.
\\
HaCa	&
The 6-parameter core system in \cite{Had_Cast}, with  transmission and recovery rates $\be_i, \ga_i>0$, a vaccination term $\al_1>0$ and a constant population with balanced birth and death rates, $\delta=\mu_i=\mu_I>0$ and $q_1=1$. 
\\
KZVH	&
The 7-parameter vaccination models of  \cite{KribsVel} adding an immunity waning rate $\al_2>0$ to the model of \cite{Had_Cast}.
\\
LiMa	&
The 8-parameter SIS-model with vaccination and varying population size of \cite{LiMa2002} keeping only 
$\ga_2=\be_2=0$ and assuming $\mu_1=\mu_2=\mu$.\protect\footnotemark\ 
\\
AABH	&
The 8-parameter SIRS-type model analyzed recently by \cite{AvramAdenane2022}, keeping only 
$\ga_1=q_2=0$ 
and  all other parameters positive.
The authors allow a varying population size by first discussing  the general case of all mortality rates being different and then concentrate on $\mu_1=\mu_2\neq \delta$ and $\Del\mu_I>0$. 
\end{longtable}
\end{flushleft}
\footnotetext{Actually the authors let $\mu_1=\mu_2=\mu=f(N)$ be a function of $N$ and put $\mu_I=\mu+\Del\mu_I$ with constant excess mortality $\Del\mu_I$. Still, $\mu=f(N)$ disappears when passing to tilde parameters \eqref{tilde_parameters}, see also Remark 
\ref{Remark}.}

\addtocounter{table}{-1} 

\begin{table}[htbp!]
\caption{Mapping models in the literature\protect\footnotemark\ 
to the present choice of parameters.  The column $\#$ counts the number of free parameters in the original models.
Passing to fractional variables $(S_1,S_2,I)$  and tilde parameters, \Eqref{tilde_parameters},
$\#_{\mathrm{eff}}$ counts the number of effectively independent parameters as determined in Eqs. \eqref{SIRS}-\eqref{AABH}.
}\label{Tab_examples}
$$
\begin{array}{l|cccccccccccc|cl}
&\al_1&\al_2&\be_1&\be_2&\ga_1&\ga_2&
\delta&\mu_1&\mu_2&\mu_I&q_1&q_2&\#&\#_{\mathrm{eff}}\\
\hline
\mathrm{Heth}&0&0&\ok&0&0&\ok&
\multicolumn{4}{c}{\delta=\mu_1=\mu_2=\mu_I}&1&0&3&3
\\\hline
\addtocounter{footnote}{1}
\mathrm{BuDr}&0&\ok&\ok&0&0&\ok&\ok& \ok&\ok&\ok&1&0&7&5 
\footnotemark\addtocounter{footnote}{-2}
\\\hline
\mathrm{SIRI_1}&0&0&\ok&\ok&0&\ok&\ok&\multicolumn{2}{c}{\mu_1=\mu_2}&\ok&1&0&6&4
\\\hline
\mathrm{SIRI_2}&0&0&\ok&\ok&\ok&0&\ok&\multicolumn{2}{c}{\mu_1=\mu_2}&\ok&0&1&6&4
\\\hline
\mathrm{SIRS}&\ok&\ok&\ok&0&\ok&\ok&
\multicolumn{4}{c}{\delta=\mu_1=\mu_2=\mu_I}&\ok&\ok&7&5
\\\hline
\mathrm{HaCa}&\ok&0&\ok&\ok&\ok&\ok&
\multicolumn{4}{c}{\delta=\mu_1=\mu_2=\mu_I}&1&0&6&6
\\\hline
\mathrm{KZVH}&\ok&\ok&\ok&\ok&\ok&\ok&
\multicolumn{4}{c}{\delta=\mu_1=\mu_2=\mu_I}&1&0&7&6
\\\hline
\mathrm{LiMa}&\ok&\ok&\ok&0&\ok&0& \ok&\multicolumn{2}{c}{\mu_i=f(N)}&\ok&\ok&\ok&8&6
\\\hline
\mathrm{AABH_1}&\ok&\ok&\ok&\ok&0&\ok&\ok&\multicolumn{2}{c}{\mu_1=\mu_2\footnotemark}&\ok&1&0&8&6
\addtocounter{footnote}{-1}
\\\hline
\mathrm{AABH_2}&\ok&\ok&\ok&\ok&\ok&0&\ok&\multicolumn{2}{c}{\mu_1=\mu_2\footnotemark}&\ok&0&1&8&6
\\\hline
\end{array}
$$
\end{table}
\addtocounter{footnote}{-1}
\footnotetext{
${\mathrm{SIRI}}$ and ${\mathrm{AABH}}$ come in two versions, since the authors also allow $\be_1<\be_2$. The subscript $1$ refers to 
$\be_1>\be_2$ and $2$ to $\be_1<\be_2$.
} 

\addtocounter{footnote}{1}
\footnotetext{\label{FN_AABH} The bulk of results in Section 5 and 6 of AABH \cite{AvramAdenane2022} assumes $\mu_1=\mu_2$.}

\addtocounter{footnote}{1}
\footnotetext{To be comparable, \Eqref{BuDr} refers to the sub-case $\mu_1=\mu_2$ in BuDr,
so $\be_2=0$ implies 
$\tilde{\be}_2=-\Del\mu_I\leq 0$. Also,
$\tilde{\ga}_1=\delta$ as in SIRI$_1$, but 
$\tilde{\al}_2=\al_2+\delta\neq\tga_1$ becomes independent.
\label{FN_BuDr}}

\addtocounter{footnote}{-1}

\noindent
Assuming $\mu_1=\mu_2$ and applying the transformations  \eqref{tilde_parameters} we get a classification in terms of the redundancy-free 6-parameter set $\A$.

\begin{align}
\Asirs	&=\A\cap\{\tilde{\beta}_2=0\}
\label{SIRS} \\
\Aheth	&=\A
\cap\{\tilde{\al}_1=0\,\land\,\tilde{\ga}_2>0\,\land\,
\tilde{\ga}_1=\tilde{\al}_2\,\land\,\tilde{\beta}_2=0\}
\label{Heth}
\\
\A_{\mathrm{SIRI}_i} &=\A
\cap\{\tilde{\al}_i\,=0\,\land\,\tilde{\ga}_j>0\,\land\,
\tilde{\ga}_i\,=\tilde{\al}_j,\,j\neq i\}
\label{SIRI}\\
\A_{\mathrm{BuDr}} &=\A
\cap\{\tilde{\al}_1=0\,\land\,\tilde{\ga}_2>0
\,\land\,\tilde{\beta}_2<0\}\footnotemark
\label{BuDr}\\
\A_{\mathrm{KZVH}} &=\A
\cap\{\tilde{\be}_2>0\}=\A_{\mathrm{HaCa}}\footnotemark
\label{KZVH}\\
\A_{\mathrm{LiMa}} &=\A
\cap\{\tilde{\be}_2<0\,\land\,\tilde{\ga}_1>0\}\footnotemark
\label{LM}\\
\A_{\mathrm{AABH}_i} &=\A\cap
\{\tilde{\ga}_j>0,\,j\neq i\}
\label{AABH}
\end{align} 
\addtocounter{footnote}{-1}

\footnotetext{For $q_1>0$ one of the three parameters 
$(\ga_1,\al_2,\delta)$ always is redundant. So 
$\A_{\mathrm{KZVH}} =\A_{\mathrm{HaCa}}$.}
\addtocounter{footnote}{1}
\footnotetext{For $q_2=1$ the mapping $(\al_1,\al_2,\ga_1,\delta)\mapsto 
(\tilde{\al}_1, \tilde{\al}_2, \tilde{\ga}_1, \tilde{\ga}_2)$ 
is bijective.}

\noindent
The dimensions of these parameter spaces are listed in the last column of Table \ref{Tab_examples}.
In summary, we arrive at

\begin{corollary}\label{Cor_Avr=JiLi+KrZa} 
Consider the dynamics of fractional variables in the models of Table \ref{Tab_examples}, for BuDr and AABH  under the restriction $\mu_1=\mu_2$.  Disregarding boundary configurations 
$\tga_i=0$ in parameter space $\A$, the following relations hold.
\begin{itemize}

\item[i)]
The model of AABH \cite{AvramAdenane2022} is isomorphic to the master model \eqref{frac_dynamics} and covers all other models.

\item[ii)]
The SIS-type model of LiMa \cite{LiMa2002} with time dependent population size coincides with the subcase 
$\min\{\be_1,\be_2\}<\Del\mu_I$ of AABH \cite{AvramAdenane2022}. 

\item[iii)]
The constant population model of KZVH \cite{KribsVel} coincides with the subcase 
$\min\{\be_1,\be_2\}>\Del\mu_I$ of AABH \cite{AvramAdenane2022}.

\item[iv)]
The subcase 
$\min\{\be_1,\be_2\}=\Del\mu_I$ of AABH \cite{AvramAdenane2022} reduces to the SI(R)S model \eqref{SIRS}.

\item[v)]
The models of HaCa \cite{Had_Cast} and KZVH  \cite{KribsVel} are isomorphic.

\end{itemize}
\end{corollary}

\section{Summary\label{Sec_summary}}
We have seen in Lemma \ref{Lem_Q} that in a large class of homogeneous compartment models with constant per capita demographic rates and time dependent total population $N$ the dynamics of fractional variables $y=Y/N$ can be rewritten such that all demographic parameters become redundant.
This way various prominent SI(R)S-type models with standard incidence, demographic parameters and possibly susceptible $R$-compartments may be normalized such that the dynamics of fractional variables appears as sub-case of a master model with zero birth and death rates, see Eqs. \eqref{SIRS}-\eqref{AABH}. 
Since apparently none of the original papers has used the identity \eqref{Q} of Lemma \ref{Lem_Q}, these relations have not been realized before. The price to pay is that in the normalized master model infection transmission rates $\tilde{\be}_i$ may also be negative. As a particular example, recent results on backward bifurcation in models with time varying total population $N(t)$,
coinciding mortality rates $\mu_1=\mu_2$ and an excess mortality  $0<\Del\mu_I<\min\{\be_1,\be_2\}$ by AABH \cite{AvramAdenane2022} are already covered by the isomorphic model with constant population of KZVH \cite{KribsVel} published in 2000. The complementary case 
$\Del\mu_I>\min\{\be_1, \be_2\}$ turns out to be isomorphic to the variable population SIS model with $\be_2=0$ published by LiMa \cite{LiMa2002} in 2002.

The normalized master model \eqref{frac_dynamics} will also be the starting point of an ongoing analysis of symmetry operations in these kinds of models giving rise to further parameter reductions, see work in progress in \cite{Nill_Symm1, Nill_Symm2}.

\ssn
{\bf Acknowledgement}
I would like to thank Florin Avram for encouraging interest and useful discussions.

\printbibliography

\end{document}